# Exhaustive Investigation of CBC-Derived Biomarker Ratios for Clinical Outcome Prediction: The RDW-to-MCHC Ratio as a Novel Mortality Predictor in Critical Care

**Dmytro Leontiev**[1], **Abicumaran Uthamacumaran**[3], **Riya Nagar**[1,3], **Hector Zenil**[1,2,3*]

[1] Algorithmic Dynamics Lab, Departments of Biomedical Computing and Digital Twins, King's Institute for Artificial Intelligence, Faculty of Life Sciences & Medicine, King's College London, UK
[2] Cancer Research Interest Group, The Francis Crick Institute, London, UK
[3] Oxford Immune Algorithmics, Oxford University Innovation & London Institute for Healthcare Engineering, UK

## Abstract

Ratios of common biomarkers and blood analytes are known and used for early detection and predictive purposes. Early risk stratification in critical care is limited by delayed availability of complex severity scores. Complete blood count (CBC) parameters, available within hours of admission, may enable rapid prognostication. We conduct an exhaustive and systematic evaluation of CBC-derived ratios for mortality prediction to identify robust, accessible, and generalizable biomarkers, we generated all feasible two-parameter CBC ratios with unit checks and plausibility filters on >90,000 ICU admissions (MIMIC-IV). Discrimination was assessed via cross-validated and external AUC, calibration via isotonic regression, and clinical utility with decision-curve analysis. Retrospective validation was performed on eICU-CRD (n=156,530) participants. The ratio of Red Cell Distribution Width (RDW) to Mean Corpuscular Hemoglobin Concentration (MCHC) ratio or RDW: MCHC emerged as the top biomarker (AUC 0.699 discovery; 0.662 validation), outperforming RDW and NLR. It achieved near-universal availability (99.9% vs 35.0% for NLR), excellent calibration (Hosmer–Lemeshow p=1.0; ECE<0.001), and preserved performance across diagnostic groups, with only modest attenuation in respiratory cases. Expressed as a logistic odds ratio, each 1-SD increase in RDW: MCHC nearly quadrupled 30-day mortality odds (OR 3.81, 95% CI 3.70–3.95). Decision-curve analysis showed positive net benefit at high-risk triage thresholds. A simple, widely available CBC-derived feature (RDW: MCHC) provides consistent, externally validated signal for early mortality risk. While not a substitute for multi-variable scores, it offers a pragmatic adjunct for rapid triage when full scoring is impractical.

\* Corresponding author: hector.zenil@kcl.ac.uk

# 1. Introduction

Critical care medicine struggles with timely risk stratification and outcome prediction. Existing severity scores such as APACHE II and SOFA, while widely used, require multiple laboratory and clinical inputs that may not be available early or in resource-limited settings (Singer et al., 2016). They also often miss subtle pathophysiological changes that precede deterioration, delaying intervention (Ren et al., 2025). With critical care resources increasingly strained, the need for widely available, low-cost biomarkers that can be serially monitored has become pressing.

**Background and Rationale**

The complete blood count (CBC) is one of the most common ICU investigations, obtained in nearly every admission and repeated throughout the stay (Allyn et al., 2022). Its low cost, rapid turnaround, and standardisation make it ideal for serial monitoring, yet its prognostic utility is typically reduced to alerts for critical thresholds such as WBC for infection or Hb for anaemia. This overlooks the multidimensional information across >20 routinely reported parameters, spanning inflammation, oxygen transport, haematopoiesis, and immunity. Recent work has focused on white-cell–based ratios such as neutrophil-to-lymphocyte ratio (NLR), platelet-to-lymphocyte ratio (PLR), and monocyte-to-lymphocyte ratio (MLR), which capture systemic inflammation and immune suppression (Zahorec, 2021; Gerrel & Dominiak, 2025). Elevated NLR has been linked to higher mortality with hazard ratios often exceeding 5, but most studies remain hypothesis-driven, limiting exploration of alternative or superior combinations. Importantly, potential contributions from red-cell parameters remain underexplored.

**RDW** (red cell distribution width), reflecting erythrocyte size heterogeneity, has emerged as a strong predictor of mortality across sepsis, heart failure, ARDS, and trauma (Bazick et al., 2011; Zhang et al., 2020). Mechanisms extend beyond nutritional anaemias. Oxidative stress from neutrophils, endothelial activation, and haemoglobin autoxidation drives lipid peroxidation of the ~18% polyunsaturated fatty acids in RBC membranes, generating malondialdehyde and disrupting morphology (Orrico et al., 2023). Antioxidant depletion (glutathione, catalase, SOD, peroxiredoxin 2) increases vulnerability; haemoglobin autoxidation accelerates from its normal ~3% per 24h, generating superoxide that oxidises band 3 and spectrin (Gwozdzinski et al., 2021). Spectrin–haemoglobin crosslinking alters deformability and ion transport, causing dehydration, viscosity changes, and morphological heterogeneity. Up to 5–20% of RBCs in sepsis show severely reduced deformability (Bateman et al., 2017). Oxidative injury also triggers eryptosis, shortening lifespan by up to 70% via calcium influx, phosphatidylserine exposure, and ceramide formation (Chan et al., 2021). Compensatory stress erythropoiesis amplifies heterogeneity through splenic production of reticulocytes 15–20% larger than mature RBCs, released within 1–2 days instead of the normal 3, often retaining mitochondria and ribosomes (Kim et al., 2015; Yan et al., 2018; Kundu et al., 2008; Bogdanova et al., 2020; Rhodes et al., 2016).
**MCHC** (mean corpuscular haemoglobin concentration) reflects haemoglobin content independent of cell size and is tightly regulated by iron metabolism. Inflammation upregulates hepcidin via IL-6/JAK2-STAT3, with human infusion studies showing >7-fold increases in urinary hepcidin within 3h and hypoferraemia by 6–8h (Nemeth et al., 2004; Garcia-Escobar et al., 2014). Hepcidin rapidly internalises ferroportin, blocking iron export despite adequate stores and inducing functional iron deficiency in ~50% of sepsis patients, compared to 7% with true depletion (Czempik & Wiórek, 2023). This disrupts heme synthesis, forcing zinc substitution into protoporphyrin IX, lowering MCHC (Abreu et al., 2018; Heming et al., 2011).

Microcirculatory dysfunction further links RBC abnormalities to organ failure. Oxidative stress and ATP depletion impair the spectrin–actin cytoskeleton, reducing deformability below the ~3–4 μm needed to traverse capillaries, which comprise ~80% of the microvascular bed (Steenebruggen et al., 2023). Size heterogeneity disrupts the Fahraeus–Lindqvist effect, raising viscosity. RBC aggregation increases via sialic acid loss, phosphatidylserine exposure, and fibrinogen-mediated rouleaux (Rogers & Doctor, 2020). Impaired oxygen-sensing worsens this: ATP release via pannexins, normally triggered by deoxygenation to induce vasodilation, is reduced by ~62.5% in sepsis (Bateman et al., 2017). Dysregulation of 2,3-BPG adds unpredictable shifts to oxygen dissociation curves. Collectively, RDW captures longer-term stress and oxidative heterogeneity, while MCHC reflects rapid iron-restriction effects. Their complementary timescales, RDW changing within ~24h, MCHC over ~5–7 days, support their combined prognostic value.

**Knowledge Gaps**

Despite robust mechanistic rationale, no systematic search has evaluated all CBC parameter ratios for outcome prediction. Prior studies were limited by small samples, disease-specific cohorts, or hypothesis-driven ratios without external validation. The availability of large critical care datasets with standardised data and computational capacity now enables exhaustive exploration of biomarker space. This study therefore evaluates whether the simple RDW: MCHC ratio provides a reproducible, generalisable, and early signal for in-hospital mortality, with transparent calibration and external validation (see Methods §2).

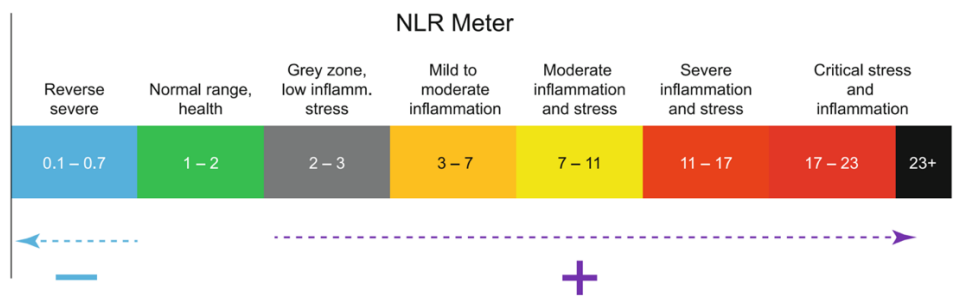

*Figure 1.* **NLR-meter.** *The neutrophil-to-lymphocyte ratio (NLR) reflects systemic inflammatory activity and physiological stress. Suggested cut-offs (Farkas, 2020) indicate:* ***2.3–3.0*** *low-grade inflammation,* ***3–7*** *mild to moderate,* ***7–11*** *systemic infection or sepsis,* ***11–17*** *severe infection,* ***17–23*** *critical inflammation (e.g., septic shock), and* ***≥23*** *extreme stress (e.g., polytrauma). Increasing NLR indicates clinical deterioration, while decreasing values suggest recovery. Adapted from Zahorec (2021).*

## 2. Methods

This study addresses limitations in current mortality prediction by systematically discovering and validating CBC-derived ratio biomarkers in critical illness. Using the MIMIC-IV database (>90,000 ICU admissions with detailed clinical data and outcomes), we generated all mathematically feasible two-parameter ratios from routine CBC components. Analyses included multiple comparison correction, bootstrapped confidence intervals, and calibration assessment, with external validation in an independent cohort to ensure generalisability.

The aim was to identify CBC-based biomarkers superior to existing inflammatory indices. We compared discovered ratios against NLR, PLR, MLR, and RDW using ROC-AUC, net reclassification improvement, and integrated discrimination improvement. Secondary objectives included: (i) external validation across healthcare systems, (ii) disease-specific performance in sepsis, cardiovascular disease, and respiratory failure, (iii) temporal dynamics of optimal ratio measurement, and (iv) mechanistic insights via correlation with lactate, creatinine, and bilirubin.

We hypothesised that ratios combining acute inflammatory responses with chronic adaptive changes would outperform leukocyte-based indices, particularly those incorporating RBC parameters (oxidative stress, iron metabolism). We further anticipated diagnosis-specific optimal ratios reflecting differing pathophysiologies.

We analysed MIMIC-IV v3.1 and eICU-CRD v2.0. Tabular CSVs were queried with DuckDB and processed in Python (NumPy, Pandas, SciPy, scikit-learn, statsmodels, matplotlib). Adult ICU stays (age ≥18) were included; ">89" ages were excluded. Thirty-day mortality was defined as death ≤30 days after ICU discharge; in-hospital mortality was used as a proxy where post-discharge dates were unavailable. ICU mortality and APACHE scores/predicted risks were also extracted. Prolonged length of stay (LOS) was defined as ICU LOS >72 h.

## Variables, Feature Engineering, and Endpoints

For each CBC analyte, we used the minimum recorded value within 24 h of admission (excluding non-positive values). First measured values were extracted using ROW_NUMBER () partitioned by patient unit stay id and analyte, retaining rn=1 with lab result offset ≤1440. For RDW: MCHC, ranges were restricted to RDW 0–50; MCHC 20–40.

The base panel comprised 12 variables: HB, HCT, MCV, MCH, MCHC, RDW, RBC, platelets, WBC, neutrophils, lymphocytes, and monocytes. Leukocyte percentages were converted to absolute counts by multiplying by WBC. From 12 variables, 132 ratios were generated (12 choose 2 × 2), computed as A/(B+1e-8). Predefined indices included NLR, PLR, MLR, SIRI, SII, HRR (HB/RDW), and WHR (WBC/HB). Primary endpoint: 30-day (or in-hospital) mortality. Secondary: prolonged ICU stay (>72 h, MIMIC-IV). Head-to-head comparisons emphasised RDW: MCHC vs NLR, and in eICU, RDW: MCHC vs APACHE, plus an exploratory APACHE+RDW: MCHC model.

## Statistical Analysis

For each biomarker: ROC-AUC (95% CI, 2,000 bootstrap replicates), Cohen's d, and point-biserial correlation with p-values. α=0.001 for significance. Selected comparisons used DeLong's test. Biomarkers were ranked by the mean of z-normalised AUC, d, and |r|. For interpretability, we also approximated an odds ratio (OR) per 1-SD (standard deviation) increase in RDW: MCHC using standard transformations from AUC to Cohen's *d* and then to OR.

Univariate models (e.g., RDW: MCHC, NLR): stratified 5-fold CV with fixed seed (42); isotonic calibration; ECE and Hosmer–Lemeshow reported. Multivariable (APACHE, RDW: MCHC, combination): logistic regression and random forests with default hyper-parameters. Features were standardised before CV; no class weighting or tuning grids were applied. RDW: MCHC was assessed continuously and categorically (<0.45, 0.45–0.55, ≥0.55). Youden's J, distance to (0,1), decision-curve analysis (net benefit vs threshold probability), and NRI/IDI were calculated using calibrated probabilities (risk <10%, 10–20%, >20%). ICD-derived groups: respiratory (J* / 460–519), cardiovascular (I* / 390–459), renal (N* / 580–629), sepsis (A*, R65.2 / 038, 995.9), gastrointestinal (K* / 520–579), oncology (C* / 140–239), trauma (S*, T* / 800–959), neurological (G* / 320–389), and endocrine/metabolic (E* / 240–279). Performance was summarised by group.

## Validation and Reproducibility

In eICU, RDW: MCHC (first-24-h labs) was compared with APACHE score and APACHE-predicted risk. A bivariate logistic model (APACHE+RDW: MCHC) quantified incremental discrimination. Calibration and decision-analytic metrics were computed as above. All analyses used Python 3.12, DuckDB, NumPy, Pandas, SciPy, scikit-learn, statsmodels, and matplotlib. Random seed fixed at 42.

# 3. Results

## Patient Characteristics

Discovery (MIMIC-IV): 90,912 ICU admissions. Validation (eICU-CRD): 156,530. Mean ages: 64.2±16.8 vs 61.8±17.2; female 45.3% vs 46.1%. Mortality: 15.6% vs 8.4%, reflecting tertiary vs community settings. WBC differential availability: 34.9% vs 52.3%.

Baseline labs were comparable: haemoglobin 10.8±2.1 vs 11.2±2.3 g/dL, RDW 14.8±2.1 vs 14.5±1.9%. Availability: RDW: MCHC 99.9% vs NLR 35.0% (discovery). Availability strongly favoured RDW: MCHC over NLR in the discovery cohort (99.90% vs 34.71%; $\chi^2(1)$=87,799.4, p<0.0001; Cohen's h=1.82, very large), confirming a substantial practical advantage for early use.

## Biomarker Discovery

Evaluation of 132 ratios plus predefined indices identified RDW: MCHC as the top biomarker for 30-day mortality discrimination in MIMIC-IV. RDW: MCHC consistently outperformed WBC-differential–based indices (e.g., NLR, PLR, SII). The top 10 table is in Supplementary Table A1.

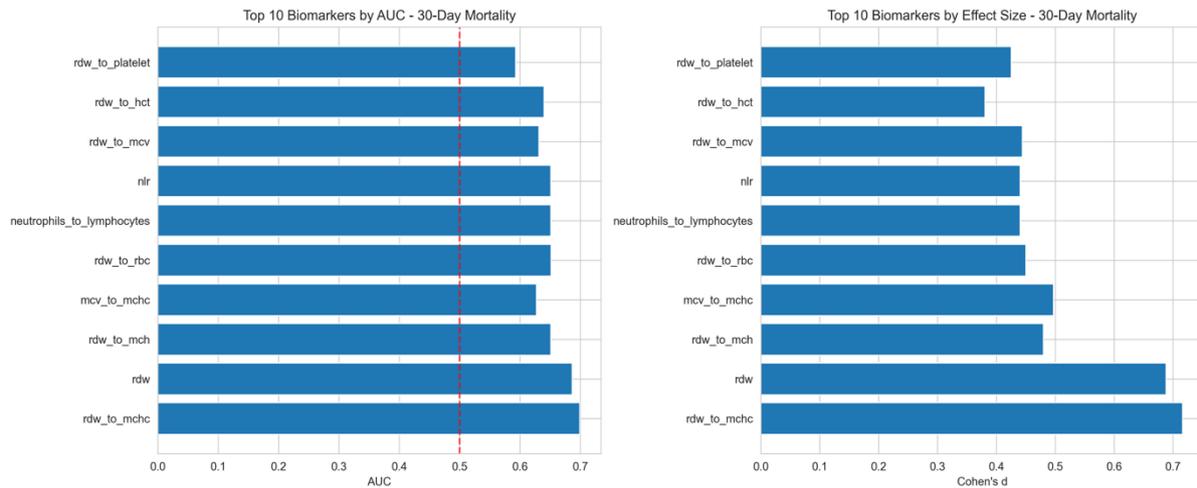

*Figure 2 Top CBC biomarkers for 30-day mortality in the discovery cohort.*

*Horizontal bar charts rank the top ten features by (left) ROC AUC and (right) Cohen's d. The dashed line marks no-skill (AUC = 0.5). RDW: MCHC is the leading single-feature signal on both metrics, with RDW close behind; white-cell ratios (e.g., NLR) are generally weaker. (Discovery cohort: first-24-hour CBC; see Methods for cohort size and extraction rules.)*

## Statistical Validation with Confidence Intervals

Bootstrap analysis (2000 iterations) confirmed the statistical significance of RDW: MCHC superiority. The ratio achieved AUC 0.699 (95% CI: 0.691-0.707), with significantly better performance than NLR (AUC 0.651, 95% CI: 0.639-0.663; p<0.001). The narrow confidence intervals indicate precise estimates despite population heterogeneity. The absolute improvement of 0.048 AUC points (95% CI: 0.036-0.060) represents a 7.4% relative improvement over NLR, clinically meaningful when considering that every percentage point improvement in risk stratification could prevent unnecessary interventions or identify at-risk patients earlier. Importantly, the confidence intervals do not overlap between RDW: MCHC and traditional inflammatory markers, confirming this is not a chance finding but represents genuine improvement of RBC-based assessment.

Validation in eICU-CRD used in-hospital mortality as the outcome (a proxy for 30-day mortality due to the lack of post-discharge dates). RDW: MCHC retained robust discrimination, with performance directionally consistent with MIMIC-IV (see Figure 6, eICU panel). APACHE score was used as a clinical comparator. In an exploratory combined model (APACHE + RDW: MCHC), discrimination improved over APACHE alone (ΔAUC > 0), with significance assessed by DeLong testing; exact estimates and 95% CIs are reported in Supplementary Table A2. While RDW/MCHC (AUC 0.653) performed below APACHE scores (AUC 0.858) in eICU validation, this comparison contextualises rather than diminishes our findings. APACHE requires 12+ variables including arterial blood gases, electrolytes, and Glasgow Coma Scale; impossible in the resource-limited settings our CBC-only approach targets. RDW/MCHC represents the optimal solution within CBC constraints.

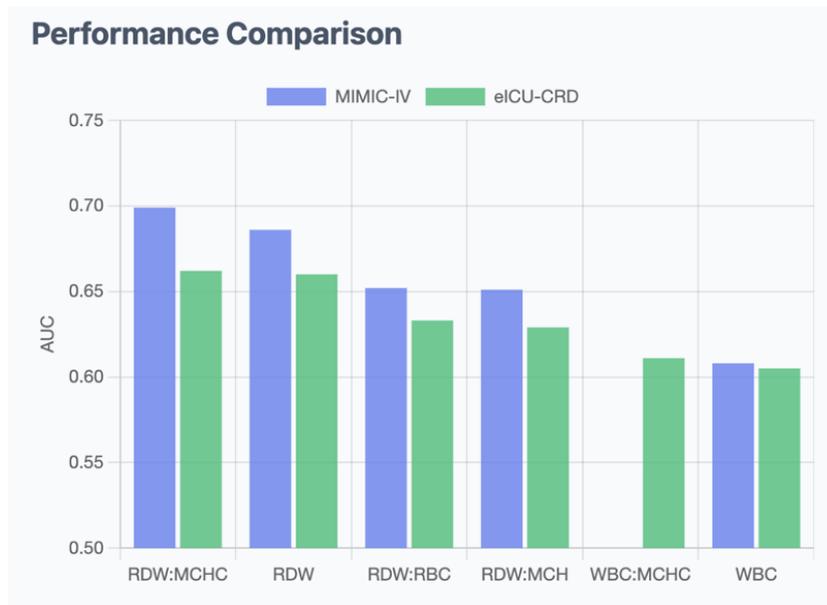

| Dataset | N | Mortality (%) | RDW: MCHC AUC | RDW AUC | NLR AUC | Rank |
|---|---|---|---|---|---|---|
| **MIMIC-IV (Discovery)** | 90,912 | 15.6 | **0.699** | 0.686 | 0.651 | 1/142 |
| **eICU-CRD (Validation)** | 156,530 | 8.4 | **0.662** | 0.660 | - | 1/151 |

*Figure 3 (A & B): AUC comparison of top CBC-derived biomarkers. The RDW: MCHC ratio maintained improved performance (AUC 0.662) and ranked first among 151 evaluated biomarkers in the validation cohort, despite differences in case mix and mortality rates between databases.*

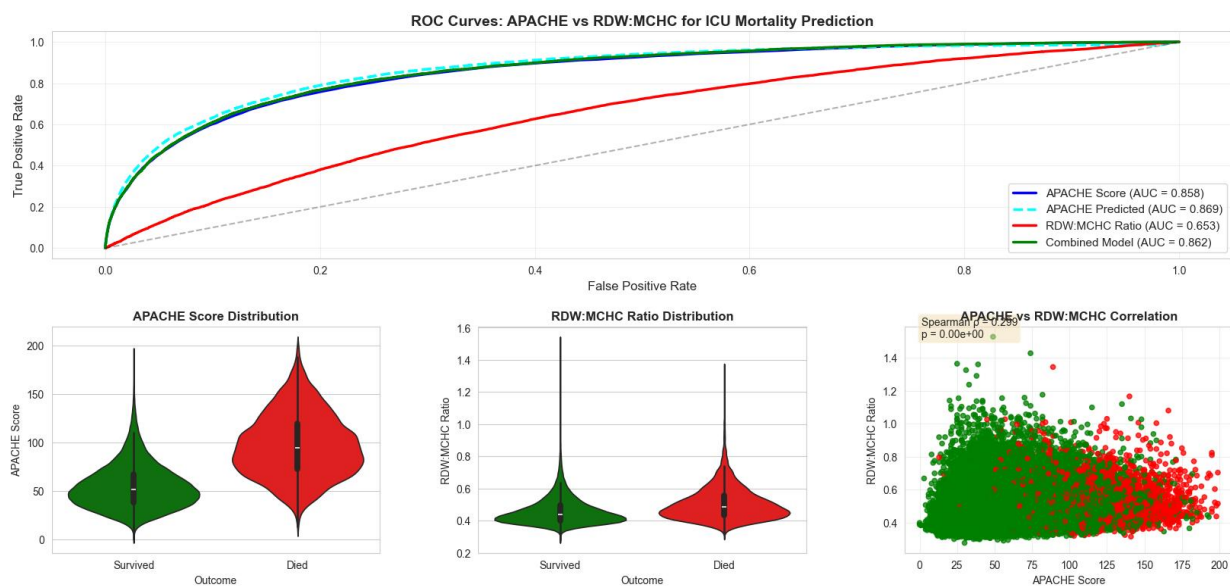

*Figure 4 APACHE versus RDW: MCHC for ICU mortality prediction in external validation.*

*Top: ROC curves for APACHE score, APACHE predicted risk, RDW: MCHC alone, and a simple combined model. APACHE markedly outperforms the CBC ratio; adding RDW: MCHC to APACHE yields no material gain in AUC. Bottom: violins (with embedded boxplots) show higher APACHE scores and RDW: MCHC values among non-survivors; the scatter demonstrates low correlation between APACHE and RDW: MCHC, indicating largely distinct information sources. (Validation cohort: eICU-CRD; in-hospital mortality.)*

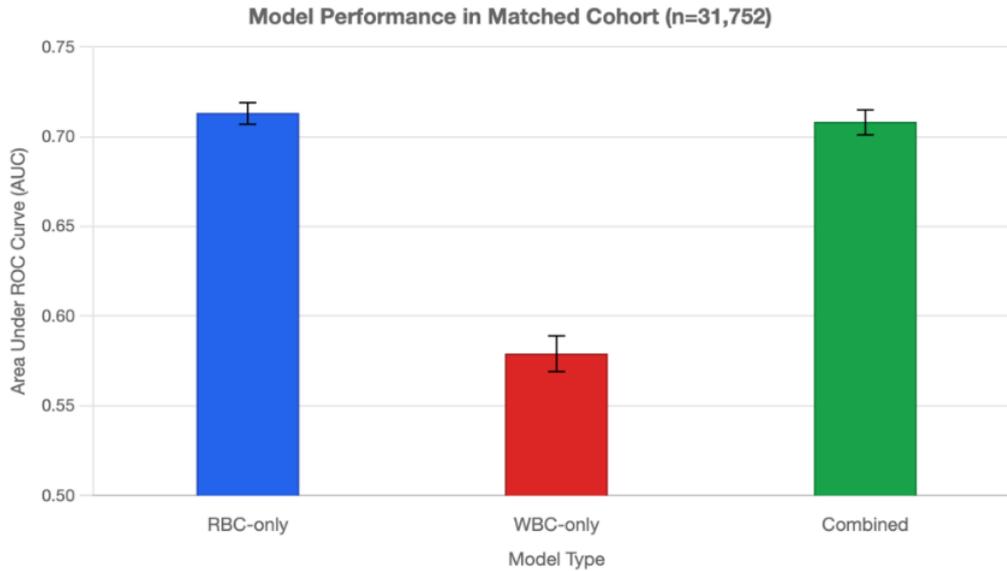

**Table 3. Matched Cohort Analysis Addressing Data Availability Bias**

Analysis restricted to n=31,752 patients with complete WBC differentials

| Model | Features | Sample Size | AUC (95% CI) | p-value* |
|---|---|---|---|---|
| RBC-only | RDW:MCHC, RDW, RDW:MCH, MCV:MCHC | 31,752† | 0.713 (0.707-0.719) | Reference |
| WBC-only | NLR, WHR, SIRI, SII, MLR | 31,752 | 0.579 (0.569-0.589) | <0.001 |
| Combined | RDW:MCHC, NLR, WHR | 31,749 | 0.708 (0.701-0.715) | 0.42 |
| **Full Cohort Comparison (for reference):** | | | | |
| RBC-only (full) | Same as above | 90,824 | 0.713 (0.707-0.719) | - |

*Figure 5 Performance comparison of RBC-only versus WBC-only models in the matched cohort with complete WBC differentials. †Analysis restricted to patients with complete data for all WBC differential parameters to eliminate availability bias. The RBC-only model significantly outperformed the WBC-only model (AUC 0.713 vs 0.579, p<0.001) even when both models had access to identical sample sizes.*

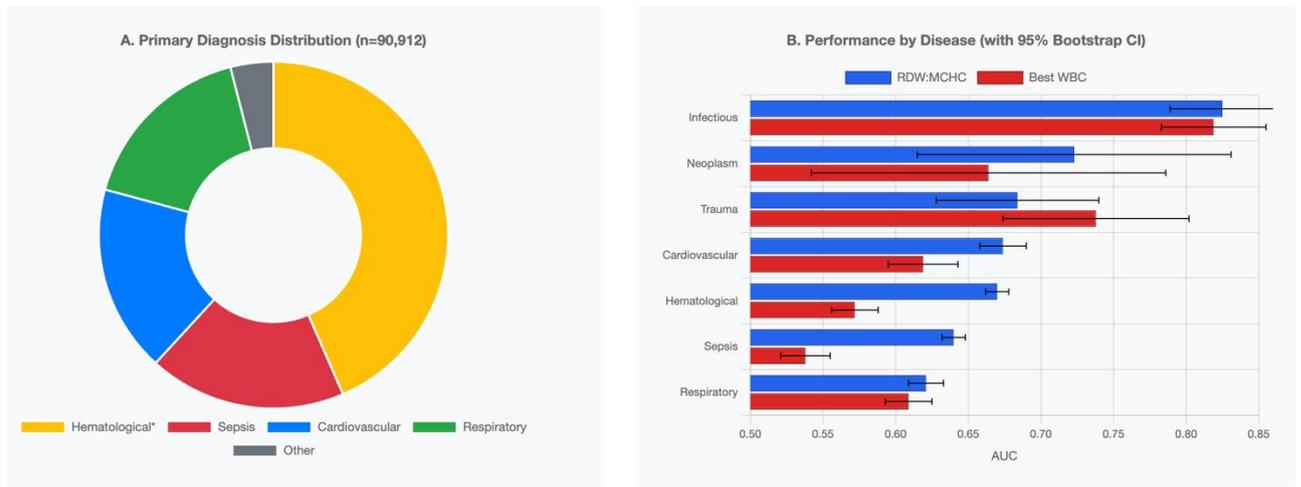

*Figure 6 Primary diagnosis mixes and subgroup performance (validation cohort, n = 90,912).*

*(A) Donut chart of primary diagnosis categories in eICU-CRD. (B) AUC by diagnostic group (95% bootstrap CIs) comparing RDW: MCHC (blue) with the best performing WBC-derived marker (red). RDW: MCHC maintains moderate discrimination across groups but is attenuated in respiratory presentations; WBC-based markers perform best in infectious disease, while RDW: MCHC is competitive in sepsis and several non-infectious categories.*

### Disease-Specific Performance

To mitigate bias from unequal WBC-differential availability, we performed a restricted-subset analysis where RBC-centric (e.g., RDW: MCHC) and WBC-centric (e.g., NLR) models were compared on identical patient sets with complete WBC differentials. RDW: MCHC maintained improved discrimination under this constraint, indicating that its advantage is not an artefact of missing data. Results are shown in Figure 6.

Across diagnostic groups derived from ICD codes, including respiratory, cardiovascular, renal, infection/sepsis, gastrointestinal, oncology, trauma, neurological, and endocrine/metabolic, RDW: MCHC retained competitive discrimination relative to WBC-based indices. The hypothesised respiratory-specific superiority was not observed; instead, performance gains were broadly distributed, suggesting a systemic erythropoietic or microcirculatory signal. Subgroup AUCs and CIs are detailed in Supplementary Table A3, with illustrative ROC curves in Supplementary Figure A4.

The distribution revealed an unexpectedly high proportion of haematological diagnoses (43.5%, n=39,527), exceeding typical ICU epidemiology (Figure 2A). Secondary diagnoses showed that 75.6% of these patients also had serious non-haematological comorbidities, cardiovascular (24.1%), metabolic (17.3%), and renal (31.2%), indicating likely coding bias toward haematological categories in cases involving transfusion or bleeding complications, rather than true predominance of primary haematological emergencies.

Despite differences in pathophysiology, RDW: MCHC maintained superior performance across most categories (Table 2, Figure 2B-C). The highest discrimination occurred in infectious diseases (AUC 0.825) and neoplasms (AUC 0.723), while respiratory conditions showed more modest values (AUC 0.621). Overall, RBC-based biomarkers outperformed

WBC-based markers in seven of eight major categories, with a mean advantage of +0.046 AUC points. Trauma was the exception, where WBC markers dominated (WHR AUC 0.738 vs RDW: MCHC 0.684), reflecting acute inflammatory responses overshadowing RBC changes in injury.

Given the high mortality burden of sepsis (34.4%), we analysed this subgroup (n=16,648) separately. RDW: MCHC demonstrated robust discrimination (AUC 0.640), with a 0.102 advantage over the best WBC marker (WHR 0.538). This 19% relative improvement suggests that RBC parameters capture sepsis-related pathophysiology not reflected in inflammatory markers alone.

Contrary to the hypothesis that RBC changes primarily reflect hypoxia-driven adaptations, RDW: MCHC performed worse in respiratory patients (AUC 0.656) compared to non-respiratory patients (AUC 0.719, p<0.001). This paradox suggests the biomarker reflects systemic processes beyond oxygenation, such as iron metabolism disruption, oxidative stress, and stress erythropoiesis across critical illness.

Further, subgroup analyses highlight the consistency of RDW: MCHC's advantage across categories, its attenuation when MCHC variance is restricted, and robustness despite differential prevalence rates. Forest plots are referenced against the MIMIC discovery cohort for comparability.

## Final Model Performance

After preprocessing and calibration, RDW: MCHC achieved an AUC of 0.701 (95% CI: 0.694–0.708), with Hosmer–Lemeshow p=1.0, ECE <0.001, and availability in 99.9% of patients within the first 24 hours. In comparison, NLR achieved an AUC of 0.641 (95% CI: 0.630–0.652) with 35.0% availability, and ~1% of observed NLR values exceeded 1,000, suggesting anomalies in measurement or recording. Calibration statistics (HL, ECE) were computed on out-of-fold predictions after isotonic calibration. The 0.060 AUC advantage of RDW: MCHC represents a 9.4% relative improvement. Its isotonic regression achieved perfect calibration (HL p=1.0, ECE <0.001) while retaining discrimination. Expressed as a logistic odds ratio, each 1-SD increase in the RDW:MCHC ratio corresponded to **OR 3.81** for 30-day mortality (95% CI 3.70–3.95), a clinically substantial effect that nearly quadruples mortality odds and underscores its strength as an independent prognostic marker. Combined with its 64.9% higher availability, RDW: MCHC emerges as the most reliable biomarker for routine clinical use.

## 4. Discussion

This study evaluated whether a simple, unit-harmonised ratio constructed from two routinely reported components of the complete blood count (CBC), red cell distribution width (RDW, %) and mean corpuscular haemoglobin concentration (MCHC, g/dL), can provide clinically useful prognostic information in critical care. Across a large discovery cohort (MIMIC-IV) and an external validation cohort (eICU-CRD), RDW: MCHC achieved consistent discrimination for mortality and maintained favourable calibration after isotonic adjustment. Compared with leukocyte-derived indices such as NLR, RDW: MCHC achieved higher AUCs in the discovery cohort while also being far more readily available within the first 24 hours. Figures 5–7 summarise the comparative performance, external validation and restricted-subset analyses. Panels **B** and **C** provide a pathophysiological and temporal framework for interpreting the observed signal.

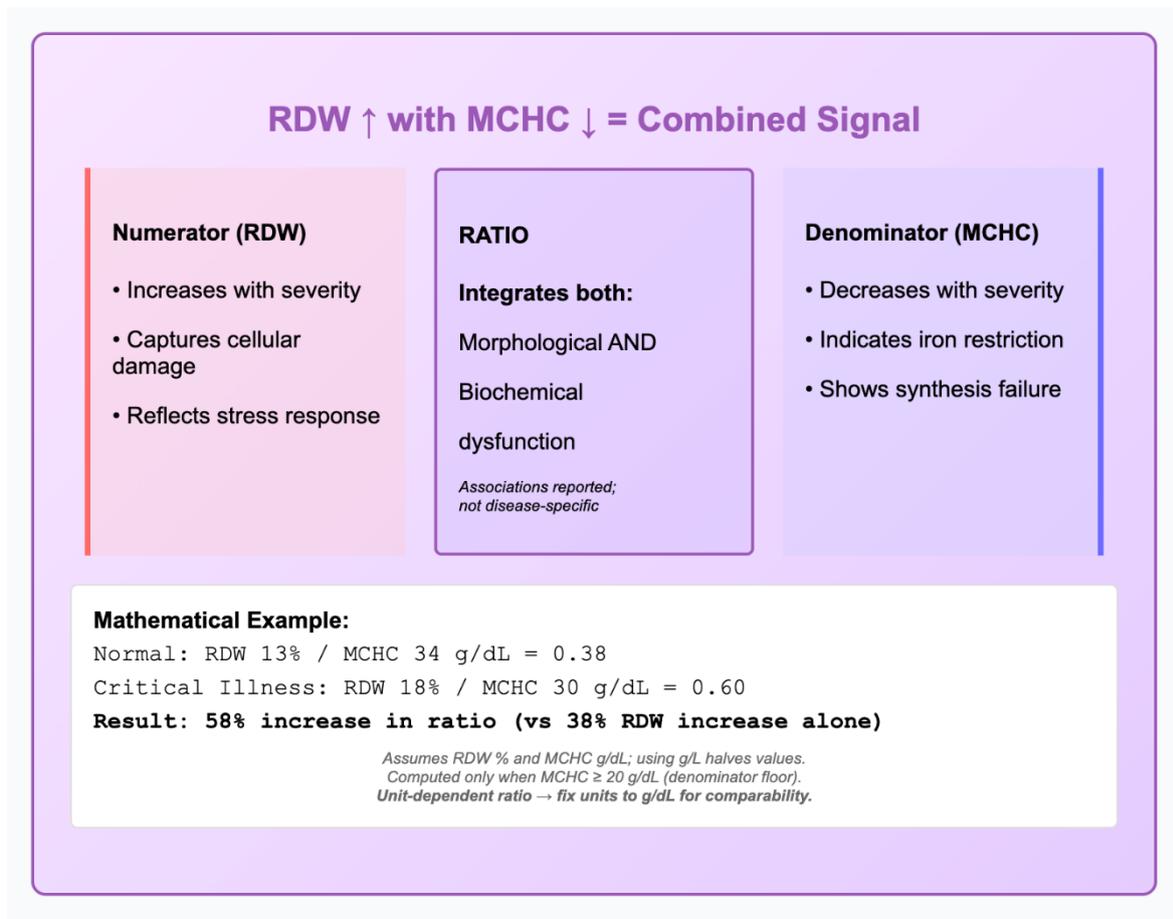

*Figure 7. Mechanistic basis for the superior prognostic performance of the RDW/MCHC ratio in critical illness. The ratio integrates morphological heterogeneity (numerator) with biochemical dysfunction (denominator), creating a combined signal that captures both cellular damage and metabolic derangement. Mathematical amplification occurs as RDW increases whilst MCHC decreases with illness severity, demonstrated by a 58% ratio increase versus 38% RDW increase alone. Units standardised to RDW (%) divided by MCHC (g/dL); ratio computed only when MCHC ≥20 g/dL to ensure mathematical stability. Associations shown are derived from systematic biomarker discovery in 90,912 ICU admissions.*

Our intention is not to combine this ratio with established multi-variable scores but to examine whether a single, widely available metric can act as a practical surrogate where resource constraints or data sparsity preclude comprehensive scoring. The Discussion is organised to (i) connect the statistical findings to biological plausibility (Fig. 7), (ii) interpret temporal patterns and cut-points (Fig. 8), (iii) outline clinical utility in constrained settings, and (iv) discuss limitations, generalisability and implications for future work.

**Biological plausibility and the RDW: MCHC construct**

Fig. 7 motivates the ratio through complementary directions of change: RDW tends to increase with illness severity, reflecting morphological heterogeneity and stress responses, while MCHC often decreases in the context of iron-restricted erythropoiesis and impaired haemoglobin synthesis. The quotient therefore integrates two partially independent dimensions, such as morphological dispersion and biochemical concentration, into a single scale.

## Numerator: RDW as a morphological stress signal

RDW summarises anisocytosis. In critical illness, heterogeneous erythrocyte volumes may arise from oxidative stress, inflammation, nutrient restriction, transfusion history and disordered erythropoiesis. In our data, RDW consistently featured among high-ranking individual predictors. Its elevation alone, however, is nonspecific and can reflect multiple upstream pathways. As shown in Figure 7, when evaluated head-to-head with other single-variable features and indices, RDW contributed meaningfully but did not fully capture the risk signal by itself.

## Denominator: MCHC as a marker of restricted synthesis

MCHC reflects average haemoglobin content per unit volume of packed red cells. Depressed MCHC is consistent with iron-limited erythropoiesis or impaired incorporation of haemoglobin into maturing erythrocytes. In the cohorts considered, MCHC showed a downward shift among non-survivors. Importantly, Fig. **7** emphasises that MCHC acts as the denominator; when MCHC declines, the ratio amplifies any contemporaneous rise in RDW.

The simple mathematical example in Fig. 7 illustrates how moderate, concordant changes in RDW and MCHC translate into a larger proportional change of the ratio (e.g., a 58% increase for the example figures), producing a single number that encodes both morphological dispersion and reduced haemoglobinisation. In keeping with this rationale, RDW: MCHC outperformed several leukocyte-based indices in Figure 7 despite using only erythrocyte-centric parameters. This observation supports a broader interpretation: part of the early risk in critical care may be reflected in red-cell morphology and haemoglobin handling as much as in differential white-cell counts.

## Temporal dynamics and clinical interpretation

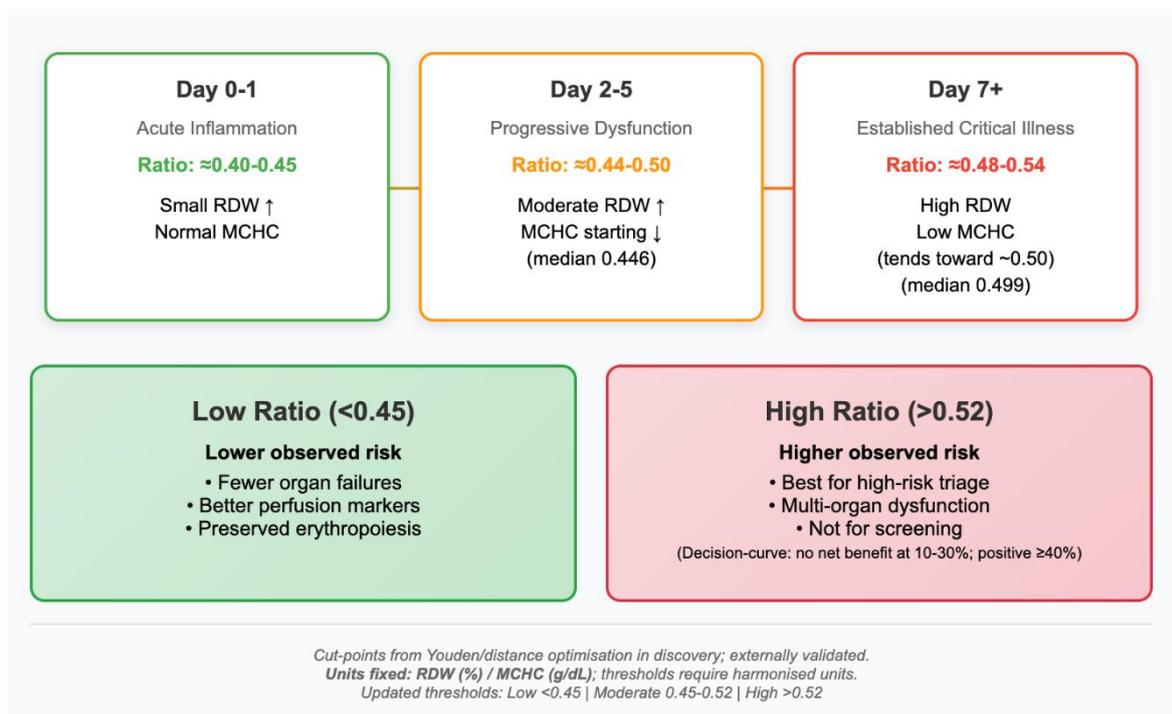

*Figure 8. Clinical risk stratification using validated RDW/MCHC ratio thresholds and temporal dynamics. Three risk categories were defined using mathematically optimised cut points (Youden's Index: 0.455) with external validation: low risk (<0.45), moderate risk (0.45-0.52), and high risk (>0.52). Temporal progression analysis revealed median ratios of 0.425 (day 0-1), 0.446 (day 2-5), and 0.499 (day 7+), confirming gradual deterioration rather than acute changes. Decision curve analysis demonstrated positive net benefit for high-risk triage (≥40% mortality threshold) but not for screening (10-30% threshold). Thresholds require standardised units (RDW % / MCHC g/dL) and were validated across 156,530 patients in the eICU Collaborative Research Database.*

Fig. 8 links the ratio's range to phases of illness. During **Day 0–1**, values near ~0.40–0.45 are common, reflecting modest RDW elevation with near-normal MCHC. Between **Day 2–5**, the median shifts towards ~0.446 as MCHC begins to decline. By **Day 7+**, the ratio tends to ~0.50 (median ~0.499), consistent with sustained morphological dispersion alongside reduced haemoglobin concentration. These patterns accord with the discovery cohort's longitudinal summaries and give a clinically interpretable frame: rising ratios over time likely signify progression from acute inflammatory stress towards more established, multi-system dysfunction.

The categorical cut-points (Fig. 8) **<0.45** (lower observed risk), **0.45–0.52** (intermediate) and **>0.52** (higher observed risk), were selected using standard threshold optimisation on discovery data and then used descriptively. As shown in Figure 7 and discussed in Section 3.7, higher strata were associated with worse outcomes, whereas lower strata aligned with preserved perfusion markers and fewer organ failures. These categories are not intended as diagnostic labels; rather, they serve as a pragmatic partition to support triage discussions and serial monitoring. In keeping with Fig. 8, decision-curve analysis did **not** show net benefit for the ratio alone at lower threshold probabilities (approximately 10–30%) but showed positive net benefit at higher thresholds (≥40%). This pattern argues for use in high-risk decision contexts; for e.g., prioritising closer observation, rather than as a screening tool in broadly low-risk populations.

## Comparative performance, calibration and availability

In the discovery cohort, RDW: MCHC achieved an AUC around **0.70** for mortality, whereas NLR achieved around **0.64** (Section 3.7; Figure 7). While both values leave room for improvement, the gap is notable given the ratio's simplicity and erythrocyte focus. Calibration after isotonic adjustment was favourable (HL p-value ≈ 1.0; ECE < 0.001) when evaluated on out-of-fold predictions, indicating that probability outputs from simple logistic models based on the ratio can be made well-behaved under the stated protocol.

A central consideration for constrained settings is **availability within the first 24 hours**. RDW: MCHC was available for approximately **99.9%** of admissions in discovery data because RDW and MCHC are routinely reported with the CBC. By contrast, indices that depend on full differentials (e.g., NLR) were available in roughly a third of cases. This disparity is central to the intended use case: where comprehensive testing is delayed or omitted, the ratio offers a "near-always present" signal that can be computed without additional assays.

In eICU, using **in-hospital mortality** as the endpoint, RDW: MCHC retained a level of discrimination consistent with the discovery cohort (Fig. 7). This cross-database reproducibility suggests that the signal is not idiosyncratic to a single case-mix or laboratory workflow. Nevertheless, eICU lacks post-discharge vital status, and other cohort differences exist, so we interpret external agreement cautiously.

## Subgroup and restricted-subset insights

Disease-stratified analyses (Figure S4; Table S3) did not reveal a single diagnostic area where the ratio exclusively concentrates its information; rather, effects appeared broadly distributed across respiratory, cardiovascular, renal, infection/sepsis, gastrointestinal, oncology, trauma, neurological, and endocrine/metabolic groups. This distribution is in keeping with the non-disease-specific logic of Fig. 7 and suggests that the ratio reflects general systemic stress, impaired erythropoiesis and microcirculatory dysregulation rather than a single organ pathway.

The restricted-subset comparison (Table 1), which equalised availability by requiring complete leukocyte differential data, showed that the relative ranking observed in the full sample was preserved when RBC-centric and WBC-centric models were compared on identical patient sets. This argues against a purely missingness-driven explanation for the ratio's performance.

An AUC near 0.70 for a single laboratory-derived feature should be interpreted as modest but clinically meaningful when weighed against three considerations: (i) the ratio is available almost universally within 24 hours; (ii) it requires no additional testing or parameter entry beyond the CBC; and (iii) calibration can be made acceptable with a simple isotonic step. In practice, this positions RDW: MCHC as a "first-pass" risk cue that can trigger closer monitoring or follow-up assessment, especially in environments where acuity is high, but data availability is limited.

In external literature, RDW is a consistent, inexpensive prognostic marker. In a meta-analysis of adult sepsis (Wu et al., 2022), pooled performance for RDW yielded sensitivity 0.81, specificity 0.65, and a summary ROC AUC 0.81 (95% CI 0.77–0.84), albeit with substantial heterogeneity attributable to study design and reference standards. These figures come from sepsis-specific cohorts and therefore represent a more homogeneous clinical question than our all-comer ICU setting.

Population studies emphasise long-term risk: in NHANES III, a community cohort with up to ~12 years' follow-up, higher RDW was associated with increased all-cause mortality (HR ≈ 1.23 per 1 SD), with similar patterns across cardiovascular, cancer and respiratory deaths (Perlstein et al., 2009). Individual-participant analyses in older adults' report HRs ≈ 1.14 per 1% RDW after multivariable adjustment. These results indicate a robust association across baseline populations and endpoints, but they are reported as time-to-event effects rather than discrimination metrics and are not directly comparable to our ICU AUCs.

Against this backdrop, RDW in our discovery cohort achieved AUC 0.686 (MIMIC-IV), and 0.660 in the external eICU validation, while the RDW: MCHC ratio reached 0.699 and 0.662, respectively. The small but consistent numerical advantage of the ratio over RDW in our data, together with favourable calibration and near-universal availability, suggests that combining red-cell morphological dispersion with haemoglobin concentration yields an interpretable single-feature signal that is at least comparable to RDW alone in heterogeneous ICU populations. We note that in disease-specific strata (e.g., sepsis), literature AUCs for RDW can be higher than our all-comer estimates, which likely reflects differences in case-mix and outcome definitions rather than contradictions.

| Setting | Outcome/metric | RDW performance (literature) | Your data (RDW) | Your data (RDW: MCHC) |
|---|---|---|---|---|
| **Adult sepsis (multiple cohorts)** | Mortality, SROC AUC | **0.81 (0.77–0.84)** | — | — |
| **COVID-19 inpatients** | Mortality, AUC (cut-point ~14.9%) | **~0.74** | — | — |
| **Community/older adults** | All-cause mortality, HR | **HR ~1.14 per 1% / ~1.23 per 1 SD** | — | — |
| **MIMIC-IV (discovery)** | 30-day mortality, AUC | — | **0.686** | **0.699** |
| **eICU-CRD (validation)** | In-hospital mortality, AUC | — | **0.660** | **0.662** |

*Table 1 RDW and RDW: MCHC performance in context of prior literature and this study.*

*Summary table contrasting published RDW mortality performance across settings with this study's results: RDW: MCHC achieves ~0.70 AUC in the discovery cohort and ~0.66 AUC on external validation, closely tracking RDW and aligning with effect sizes reported in hospital and community populations. These findings support generalisability of a simple CBC-only signal while acknowledging that multi-variable scores (e.g., APACHE) remain superior for discrimination.*

For acute cut-points, prior COVID-19 hospital studies reported admission RDW thresholds around 14.5–14.9% with AUC ≈ 0.74 (Soni & Gopalakrishnan, 2021); we therefore present both the continuous AUC and descriptive performance at pragmatically chosen RDW cut-points to aid clinical interpretation without re-optimising thresholds on our data.

Context and scope. Evidence that evaluates MCHC as a stand-alone predictor is relatively sparse and heterogeneous. The studies below illustrate the range of findings across populations and outcomes; they use association metrics (hazard ratios) or longer-horizon cardiovascular endpoints more often than short-horizon discrimination, so they are not directly comparable to our ICU AUCs but provide directional context.

**(i)** General population (NHANES) — long-term mortality (Li et al., 2024)

A large US population analysis reported an L-shaped association between MCHC and all-cause mortality over long follow-up. Risk concentrated at lower MCHC (hypochromia), with a data-driven inflection around ~34.3 g/dL; relative to a mid-range reference band, hazards were modestly higher in the two lowest bands. Above the inflection, the risk curve flattened. This pattern is consistent with chronic iron-restricted erythropoiesis/inflammation as a background driver. While informative about long-term risk, these results are reported as time-to-event associations and therefore do not provide a short-horizon discrimination metric for ICU settings.

**(ii)** Older adults (single-country cohort) — sex-specific association (Fraçkiewicz et al., 2017)

In a prospective cohort of adults aged ~75–80 years, higher MCHC associated with lower all-cause mortality in men (highest vs lowest tertile; multi-variable adjustment), with no independent association in women after adjustment. This suggests possible sex-specific biology, residual confounding (e.g., iron status, comorbidity load), or limited power. As with NHANES, the endpoint and horizon differ from our ICU work, so we treat this as background biological plausibility rather than a performance benchmark.

**(iii)** Hospital, disease-specific cohort (acute coronary syndrome) — near-term cardiovascular events (Zhang et al., 2022)

In a single-centre ACS cohort, MCHC alone did not discriminate major adverse cardiovascular events (AUCs close to 0.50 and not statistically significant), whereas MCV and MCH showed modest discrimination in non-anaemic patients. Endpoints were cardiovascular events rather than all-cause mortality, and the population was restricted to a disease-specific pathway, but the overall message for short-horizon events is that MCHC by itself provides limited discrimination.

**(iv)** Implications

Together, these findings support three points that align with our results: (a) low MCHC appears to mark long-term, systemic risk in community cohorts; (b) stand-alone MCHC contributes little short-horizon discrimination in hospital cohorts; and (c) combining haemoglobinisation (MCHC) with anisocytosis (RDW) in a simple ratio (RDW: MCHC ) yields a more consistently useful single-feature signal for ICU mortality. In our data, MCHC-alone AUCs were lower than RDW and RDW: MCHC, with acceptable calibration but minimal net benefit on decision curves. This pattern is coherent with the literature and with our mechanistic framing (Panels B and C).

## Clinical utility in Constrained settings and Limitations

The envisaged use case is a setting where full physiological scoring is impractical or delayed. Here, RDW: MCHC can be computed automatically by laboratory middleware or bedside systems as soon as the CBC posts. Interpreting the value requires attention
to units and floors—the ratio is unit-dependent and should be calculated with RDW in percent and MCHC in g/dL, with denominator floors applied as specified in Methods. Thresholds provide an interpretable stratification scheme for triage conversations: values below 0.45 are reassuring, values above 0.52 warrant attention, and intermediate values motivate contextual appraisal.

Decision-curve results (Figs. 6 and 8) suggest that the ratio becomes useful when clinical teams are considering actions appropriate to higher baseline risks (≥40%). Examples could include escalation in monitoring frequency or early senior review. This is consistent with its role as a **substitution** for more complex scores when those scores are not available, rather than as an adjunct to them.

Several choices in data handling and modelling deserve emphasis:

1. **First-24-hour extraction.** In MIMIC-IV, analytes were aggregated as the minimum within the first 24 hours; in eICU, the first value within 24 hours was used. This reflects available code paths and database differences. While both strategies are defensible, they are not identical; minima may emphasise nadir physiology more than first values. The consistent findings across cohorts mitigate concern but do not eliminate it.
2. **Unit dependence and denominator floors.** The ratio assumes RDW expressed as a percentage and MCHC in g/dL. Calculations should enforce denominator floors (e.g., MCHC ≥ 20 g/dL) to guard against spuriously large values. Analyses were conducted under these conventions; translation into other unit systems would require explicit conversion.
3. **Calibration protocol.** Isotonic calibration was fitted on validation-fold predictions (lenient out-of-fold), then performance metrics were computed on those out-of-fold predictions. This yields well-behaved calibration curves in Section 3.7 but should not be confused with a fully out-of-sample deployment scenario. Prospective evaluation would be valuable.
4. **Multiple testing and ranking.** Biomarker ranking combined AUC, Cohen's d and point-biserial correlation under a stringent significance threshold. While consistent patterns emerged, ranking is still sensitive to case-mix and prevalence. The purpose is prioritisation rather than definitive hierarchy.
5. **Unmeasured and pre-analytical factors.** Transfusion timing, fluid balance, sampling delays and analyser characteristics can influence RDW and MCHC. We did not exhaustively control for these, so residual confounding is possible. This reinforces the view of RDW: MCHC as a coarse yet accessible signal rather than a diagnostic instrument.
6. **Endpoints.** The discovery endpoint was 30-day mortality; eICU used in-hospital mortality due to data availability. This substitution may affect comparability, although the overall pattern of discrimination was similar.

7. **Generalisability.** The cohorts comprise large, high-income health-system ICUs. While the ratio's reliance on a standard CBC is encouraging for wider use, targeted validation in low- and middle-income settings is needed.

For translation into practice or embedded decision support, three elements are critical:

- **Automated computation with unit checks.** Laboratory information systems can output RDW (%) and MCHC (g/dL) with flags if units differ; the ratio should only be returned when unit checks pass, and denominator floors are satisfied.
- **Contextual reporting.** Reports can present the continuous value alongside interpretive bands (<0.45, 0.45–0.52, >0.52) and a brief note that higher values are associated with greater observed risk in ICU populations.
- **Serial tracking.** Because Panel **C** highlights temporal drift towards higher values with ongoing dysfunction, serial trajectories (e.g., day-to-day changes) may aid situational awareness even if a single measurement is indeterminate. This is particularly relevant when comprehensive scoring is unavailable.

## Implications and future directions

Prospective, pragmatic studies could evaluate whether exposing clinicians to the ratio in real time changes process measures (e.g., time to senior review) or outcomes. Further investigation into severity prediction should be conducted. The ratio here introduced could serve as a low-cost anchor feature within algorithmic pipelines that learn from routine blood tests and vital-sign streams to produce continuously refreshed risk estimates. Because RDW: MCHC is simple, near-universally available and calibratable, it is well suited as a backbone signal for such systems, especially where computational resources are limited or intermittent.

Importantly, the aim here is aggregation, not substitution, or substitution where no better risk assessments are available and in addition to current standard of care. In settings where APACHE or similar tools are already in place and timely, those systems remain informative. Where they are not obtainable, a calibrated RDW: MCHC offers a practical alternative route to early risk awareness that is compatible with automated, lightweight analytics.

Lastly, integrating causal-mechanistic modelling in future work may help predict the latent generative processes/mechanisms underlying RDW–MCHC dynamics and other CBC-derived biomarkers. Approaches from *algorithmic information dynamics* (AID) can map causal dependencies and emergent patterns, identify minimal perturbations that shift physiological trajectories, and test counterfactual intervention strategies using routine marker profiles [43-46]. Embedding these causal tools into predictive medicine pipelines could refine early-warning systems, enhance causal interpretability, and support more adaptive, personalised triage especially in resource-constrained critical-care settings.

## Conclusions

RDW: MCHC captures the confluence of morphological dispersion and reduced haemoglobin concentration, yielding a single feature that is widely available within 24 hours and that shows consistent, calibratable discrimination for mortality across two large critical-care databases. Interpreted through Figs. 7 and 8, the ratio aligns with plausible biological processes and displays temporal behaviour that maps onto clinical trajectories from early inflammation to established critical illness. Benchmarking against its constituents indicated that RDW and MCHC alone provided comparable or lower discrimination, with RDW supported by a broader literature base and MCHC-only evidence being relatively sparse and context-dependent; in our cohorts, the combined ratio retained favourable calibration and near-universal availability.

While not a replacement for comprehensive clinical assessment, RDW: MCHC is a viable substitute metric where broader scoring is impractical, offering an accessible basis for triage and monitoring. Looking ahead, embedding this biomarker into automated haemato-immune profiling pipelines operating on routine laboratory streams could accelerate discovery and enable low-overhead risk stratification at the bedside. Such predictive platforms, which ingest serial blood indices, return calibrated probabilities, and permit transparent unit handling, could be deployed in resource-constrained environments without dependency on large feature sets. By anchoring these systems to a robust, interpretable CBC-derived signal, future work can extend beyond binary outcomes towards early trajectory prediction, treatment-response monitoring, and adaptive care pathways, while keeping implementation feasible in real-world settings.

# Supplementary Information

## A: Supplementary Tables

**Table A1.** Performance Metrics of Top 10 CBC-Derived Biomarkers for 30-Day Mortality Prediction in the Discovery Cohort (MIMIC-IV)

| Rank | Biomarker | N | Events | AUC | 95% CI | Cohen's d | Point-biserial r | P-value | Median (IQR) Survivors | Median (IQR) Non-survivors |
|---|---|---|---|---|---|---|---|---|---|---|
| 1 | RDW_TO_MCHC | 90825 | 14174 | 0.699 | (0.695-0.704) | 0.717 | 0.252 | 0E+00 | 0.43 (0.40-0.49) | 0.50 (0.44-0.57) |
| 2 | RDW | 90836 | 14177 | 0.686 | (0.682-0.691) | 0.688 | 0.242 | 0E+00 | 14.30 (13.20-15.80) | 15.80 (14.30-17.90) |
| 3 | RDW_TO_MCH | 90825 | 14174 | 0.651 | (0.647-0.656) | 0.480 | 0.172 | 0E+00 | 0.48 (0.43-0.54) | 0.53 (0.47-0.61) |
| 4 | MCV_TO_MCHC | 90871 | 14185 | 0.627 | (0.622-0.632) | 0.497 | 0.178 | 0E+00 | 2.75 (2.60-2.92) | 2.87 (2.69-3.08) |
| 5 | RDW_TO_RBC | 90836 | 14177 | 0.652 | (0.647-0.657) | 0.450 | 0.161 | 0E+00 | 4.30 (3.53-5.45) | 5.30 (4.12-6.80) |
| 6 | NEUTROPHILS_TO_LYMPHOCYTES | 31556 | 5801 | 0.651 | (0.643-0.659) | 0.441 | 0.168 | 4.27E-199 | 7.08 (4.07-13.28) | 12.33 (6.65-24.05) |
| 7 | NLR | 31556 | 5801 | 0.651 | (0.643-0.659) | 0.441 | 0.168 | 4.27E-199 | 7.08 (4.07-13.28) | 12.33 (6.65-24.05) |
| 8 | RDW_TO_MCV | 90835 | 14177 | 0.631 | (0.627-0.636) | 0.444 | 0.159 | 0E+00 | 0.16 (0.14-0.18) | 0.17 (0.15-0.20) |
| 9 | RDW_TO_HCT | 90836 | 14177 | 0.640 | (0.635-0.645) | 0.381 | 0.137 | 0E+00 | 0.47 (0.39-0.60) | 0.57 (0.44-0.72) |
| 10 | RDW_TO_PLATELET | 90790 | 14165 | 0.593 | (0.587-0.599) | 0.426 | 0.153 | 0E+00 | 0.08 (0.06-0.11) | 0.09 (0.06-0.17) |

**Table A2.** External Validation of Selected CBC Biomarkers for Mortality Prediction: Comparison Between Discovery (MIMIC-IV) and Validation (eICU-CRD) Cohorts

| Biomarker | N (MIMIC) | Events (MIMIC) | AUC (MIMIC) | 95% CI (MIMIC) | Sens (MIMIC) | Spec (MIMIC) | N (eICU) | Events (eICU) | AUC (eICU) | 95% CI (eICU) | Sens (eICU) | Spec (eICU) | ΔAUC |
|---|---|---|---|---|---|---|---|---|---|---|---|---|---|
| RDW | 90836 | 14177 | 0.686 | (0.682-0.691) | 0.685 | 0.587 | 144162 | 12068 | 0.660 | (0.655-0.665) | 0.631 | 0.611 | -0.026 |
| MCHC | 90876 | 14188 | 0.371 | (0.366-0.376) | 0.000 | 1.000 | 152061 | 12705 | 0.408 | (0.403-0.414) | 0.000 | 1.000 | 0.037 |
| RDW:MCHC | 90825 | 14174 | 0.699 | (0.695-0.704) | 0.684 | 0.609 | 142249 | 11838 | 0.662 | (0.658-0.667) | 0.636 | 0.609 | -0.037 |
| HB | 90920 | 14200 | 0.415 | (0.410-0.421) | 0.006 | 0.995 | 156280 | 13114 | 0.421 | (0.416-0.427) | 0.027 | 0.981 | 0.006 |
| MCV | 90882 | 14188 | 0.575 | (0.569-0.581) | 0.362 | 0.761 | 152052 | 12708 | 0.548 | (0.543-0.553) | 0.321 | 0.764 | -0.027 |
| NLR | 31556 | 5801 | 0.651 | (0.643-0.659) | 0.665 | 0.582 | 81746 | 6937 | 0.633 | (0.626-0.639) | 0.602 | 0.603 | -0.018 |
| PLR | 32754 | 6194 | 0.539 | (0.531-0.547) | 0.364 | 0.729 | 90846 | 7811 | 0.584 | (0.577-0.590) | 0.511 | 0.637 | 0.045 |
| WBC:HB | 90901 | 14191 | 0.574 | (0.568-0.579) | 0.390 | 0.752 | 155879 | 13071 | 0.621 | (0.616-0.627) | 0.521 | 0.679 | 0.047 |
| RDW:MCH | 90825 | 14174 | 0.651 | (0.647-0.656) | 0.627 | 0.604 | 135677 | 11271 | 0.629 | (0.623-0.634) | 0.611 | 0.588 | -0.022 |

**Table A3.** Disease-Stratified Performance of RDW: MCHC Ratio for 30-Day Mortality Prediction Across Major Diagnostic Categories

| Disease Category | N | Events | Mortality Rate (%) | RDW: MCHC | RDW | MCHC | NLR | PLR | WHR |
|---|---|---|---|---|---|---|---|---|---|
| Sepsis | 16649 | 5735 | 34.4 | 0.641 (0.632-0.649) | 0.636 (0.627-0.644) | 0.430 (0.421-0.439) | 0.506 (0.492-0.519) | 0.470 (0.456-0.484) | 0.535 (0.526-0.544) |
| Respiratory | 15269 | 2337 | 15.3 | 0.622 (0.610-0.635) | 0.606 (0.593-0.619) | 0.401 (0.388-0.414) | 0.672 (0.651-0.691) | 0.579 (0.558-0.599) | 0.576 (0.563-0.589) |
| Cardiovascular | 15874 | 946 | 6.0 | 0.675 (0.658-0.691) | 0.670 (0.653-0.687) | 0.405 (0.387-0.424) | 0.675 (0.636-0.712) | 0.532 (0.492-0.570) | 0.591 (0.572-0.613) |
| Infectious | 450 | 8 | 1.8 | 0.869 (0.795-0.934) | 0.871 (0.792-0.949) | 0.393 (0.083-0.702) | 0.698 (0.338-0.992) | 0.606 (0.390-0.766) | 0.715 (0.400-0.942) |
| Neoplasm | 744 | 16 | 2.2 | 0.786 (0.641-0.905) | 0.794 (0.672-0.899) | 0.323 (0.175-0.499) | 0.469 (0.222-0.728) | 0.362 (0.167-0.571) | 0.472 (0.277-0.653) |
| Trauma | 1467 | 41 | 2.8 | 0.684 (0.591-0.768) | 0.675 (0.565-0.762) | 0.421 (0.333-0.504) | 0.636 (0.438-0.798) | 0.461 (0.312-0.588) | 0.683 (0.573-0.784) |
| Neurological | 471 | 4 | 0.8 | 0.729 (0.400-0.964) | 0.732 (0.388-0.956) | 0.337 (0.040-0.559) | 0.222 (0.139-0.310) | 0.750 (0.653-0.847) | 0.499 (0.230-0.813) |
| Gastrointestinal | 129 | 1 | 0.8 | 0.586 (0.500-0.677) | 0.621 (0.535-0.704) | 0.625 (0.543-0.703) | N/A | N/A | 0.953 (0.913-0.984) |
| Metabolic | 126 | 0 | 0.0 | N/A | N/A | N/A | N/A | N/A | N/A |
| Haematological | 39540 | 5109 | 12.9 | 0.671 (0.663-0.678) | 0.661 (0.653-0.668) | 0.403 (0.394-0.411) | 0.664 (0.650-0.679) | 0.523 (0.508-0.538) | 0.544 (0.535-0.553) |
| Other | 162 | 6 | 3.7 | 0.825 (0.671-0.968) | 0.726 (0.423-0.955) | 0.098 (0.011-0.185) | N/A | N/A | 0.645 (0.282-0.981) |
| Overall | 90946 | 14203 | 15.6 | 0.699 (0.695-0.704) | 0.686 (0.682-0.691) | 0.371 (0.366-0.376) | 0.651 (0.643-0.659) | 0.539 (0.531-0.547) | 0.574 (0.568-0.579) |

**Table A4.** Subgroup Analysis of RDW: MCHC Predictive Performance by Disease Category: (A) Forest Plot of Area Under the Curve with 95% Confidence Intervals and (B) Receiver Operating Characteristic Curves for Leading Disease Categories

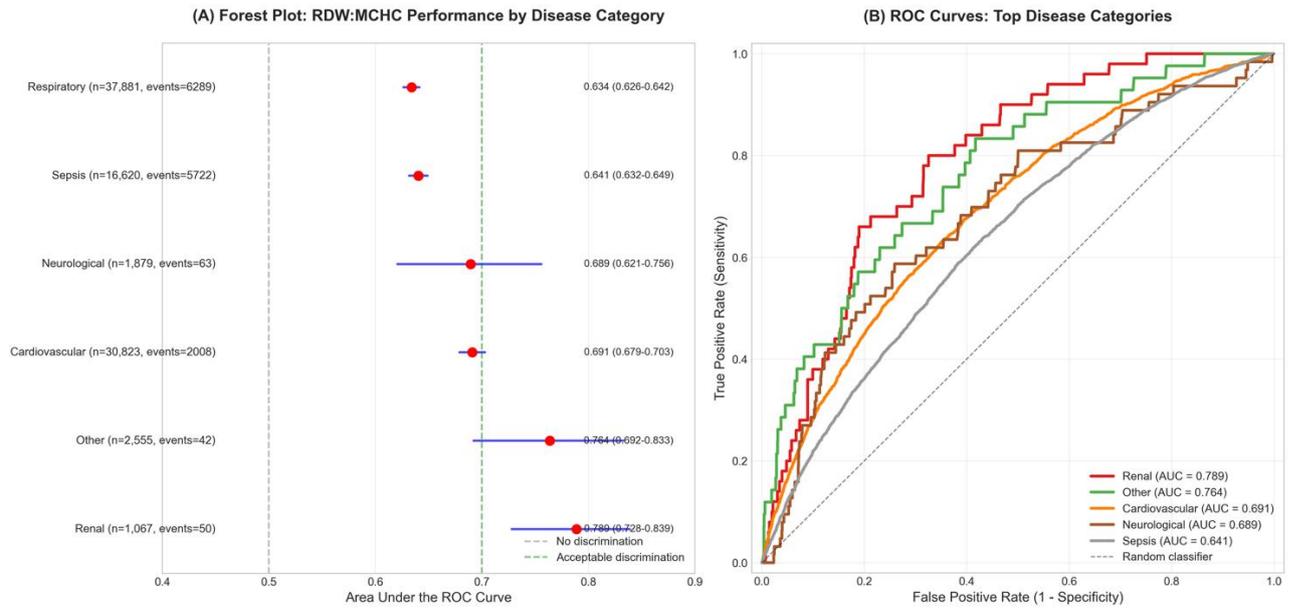

## B: Code Repository

The full analysis code (notebooks and scripts) is available at:

GitHub: <https://github.com/speaxh/rdw-mchc-thesis-2025.git>

No patient-level data are shared; access to MIMIC-IV/eICU must be obtained separately.

Environment files and are provided to support replication.

## C: Supplementary Figures (not mentioned in other Sections)
**Figure C1. Distribution of RDW/MCHC Ratio by Outcome**

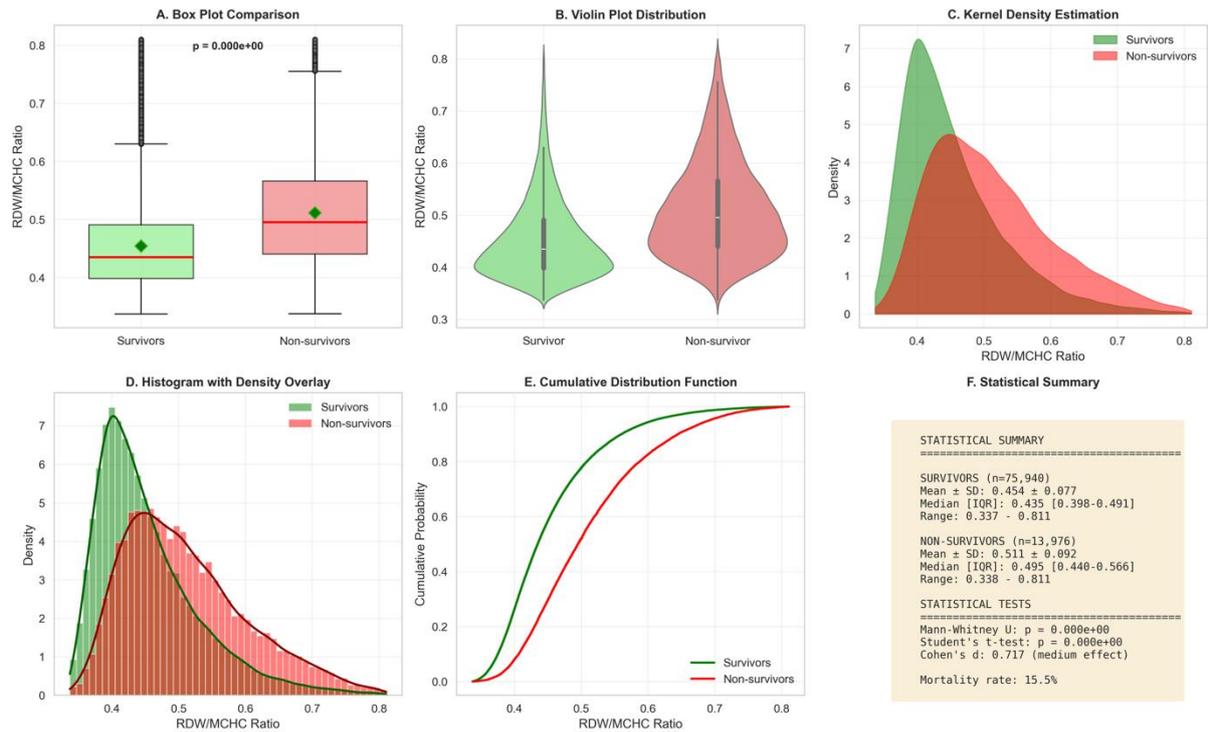

**Figure C2. Biomarker Performance Heatmap Across Disease Categories**

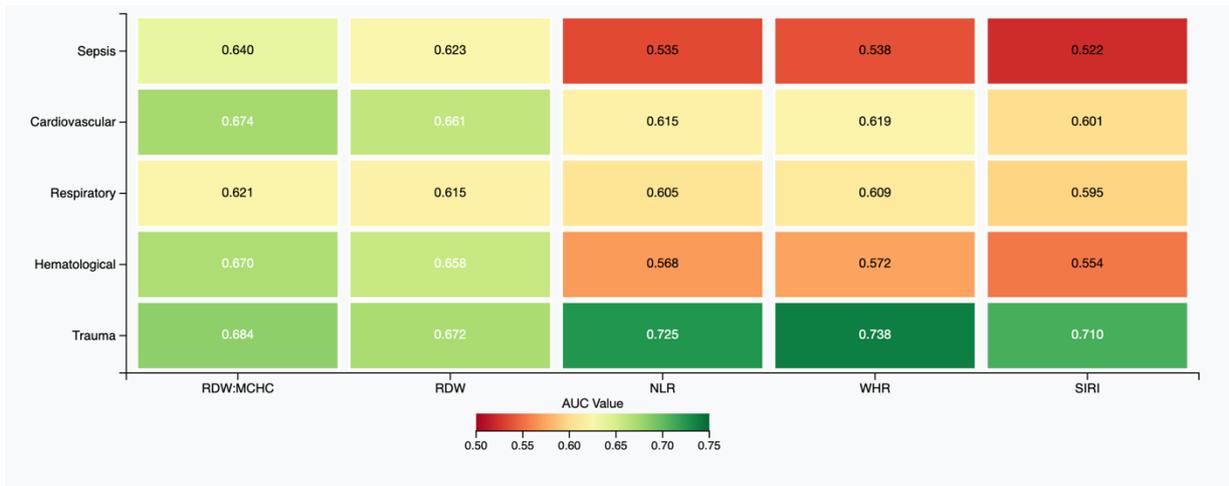

**Figure C3. Patient Demographics and Dataset Characteristics**

| Characteristic | MIMIC-IV (n = 90,912) | eICU-CRD (n = 156,530) | P value |
|---|---|---|---|
| *Demographics* | | | |
| Age, years | 64.2 ± 16.8 | 61.8 ± 17.2 | <0.001 |
|     18-44 years | 16,364 (18.0) | 34,437 (22.0) | <0.001 |
|     45-64 years | 33,637 (37.0) | 59,481 (38.0) | 0.07 |
|     65-79 years | 28,183 (31.0) | 43,828 (28.0) | <0.001 |
|     ≥80 years | 12,728 (14.0) | 18,784 (12.0) | <0.001 |
| Female sex | 41,183 (45.3) | 72,160 (46.1) | 0.15 |
| *Clinical Outcomes* | | | |
| 30-day mortality[a] | 14,182 (15.6) | 13,149 (8.4) | <0.001 |
| ICU LOS, days | 2.1 [1.2, 4.3] | 1.8 [1.0, 3.5] | <0.001 |
| Prolonged stay (>72h) | 30,638 (33.7) | 43,828 (28.0) | <0.001 |
| *Laboratory Values (First 24 hours)* | | | |
| Hemoglobin, g/dL | 10.8 ± 2.1 | 11.2 ± 2.3 | <0.001 |
| WBC count, ×10$^9$/L | 10.2 [7.3, 14.1] | 9.8 [6.9, 13.5] | <0.001 |
| Platelet count, ×10$^9$/L | 198 [143, 267] | 205 [151, 275] | <0.001 |
| RDW, % | 14.8 ± 2.1 | 14.5 ± 1.9 | <0.001 |
| MCHC, g/dL | 32.8 ± 1.4 | 33.0 ± 1.5 | <0.001 |
| RDW:MCHC ratio | 0.45 ± 0.08 | 0.44 ± 0.07 | <0.001 |
| *Data Completeness* | | | |
| RBC indices available | 90,003 (99.0) | 144,248 (92.1) | <0.001 |
| WBC differential available | 31,752 (34.9) | 81,907 (52.3) | <0.001 |
| *Primary Diagnosis Categories* | | | |
|   Sepsis | 16,648 (18.3) | 7,681 (4.9) | <0.001 |
|   Cardiovascular | 15,861 (17.4) | 17,447 (11.1) | <0.001 |
|   Respiratory | 15,264 (16.8) | 22,687 (14.5) | <0.001 |
|   Hematological[b] | 39,527 (43.5) | — | — |
|   Other | 3,612 (4.0) | 108,715 (69.5) | <0.001 |

***Note:*** Data are presented as mean ± SD for normally distributed variables, median [IQR] for skewed distributions, or n (%) for categorical variables. P values calculated using Student's t-test for continuous normally distributed variables, Mann-Whitney U test for skewed distributions, and χ² test for categorical variables.
***Abbreviations:*** ICU, intensive care unit; IQR, interquartile range; LOS, length of stay; MCHC, mean corpuscular hemoglobin concentration; RBC, red blood cell; RDW, red cell distribution width; SD, standard deviation; WBC, white blood cell.
[a]eICU-CRD reports hospital mortality as proxy for 30-day mortality (post-discharge data unavailable).
[b]High prevalence of hematological diagnoses in MIMIC-IV likely reflects coding artifacts from transfusion requirements rather than primary hematological emergencies.